\documentclass[trackchanges,twocolumn]{aastex701}

\usepackage{graphicx} 
\usepackage[nointegrals]{wasysym} 
\usepackage{graphicx, amsmath, amssymb}
\usepackage{booktabs}
\usepackage{rotating}

\newcommand{\citeg}[1]{\citep[e.g.,][]{#1}}

\begin{document}
\title{Star Formation and Nebular Attenuation from Pa$\alpha$, Br$\alpha$, and Br$\beta$ in Massive Dust-obscured Galaxies at Cosmic Noon using JWST}

\author[orcid=0009-0001-4693-1519,gname=Virginia, sname=Vanicek]{Virginia Vanicek}
\affiliation{Department of Astronomy, The University of Texas at Austin, 2515 Speedway Boulevard, Austin, TX 78712, USA}
\affiliation{Cosmic Frontier Center, The University of Texas at Austin, Austin, TX 78712}
\affiliation{Department of Astronomy, University of Maryland, College Park, MD 20742, USA}
\email{vvanicek@umd.edu}  

\author[orcid=0000-0002-6149-8178,gname=McKinney,sname=Jed]{Jed McKinney}
\altaffiliation{NASA Hubble Fellow}
\affiliation{Department of Astronomy, The University of Texas at Austin, 2515 Speedway Boulevard, Austin, TX 78712, USA}
\affiliation{Cosmic Frontier Center, The University of Texas at Austin, Austin, TX 78712}
\email{}

\author[0000-0001-8592-2706]{Alexandra Pope} 
\affiliation{Department of Astronomy, University of Massachusetts, Amherst, MA 01003, USA}
\email{}

\author[0000-0002-1917-1200]{Anna Sajina}
\affiliation{Department of Physics and Astronomy, Tufts University, Medford, MA 02155, USA}
\email{}

\author[0000-0002-8909-8782]{Stacey Alberts} 
\affiliation{AURA for the European Space Agency (ESA), Space Telescope Science Institute, 3700 San Martin Dr., Baltimore, MD 21218, USA} \email{salberts@stsci.edu}
\email{}

\author[orcid=0000-0002-9720-3255, gname=Stone,sname=Meredith] {Meredith Stone}
\affiliation{Steward Observatory, University of Arizona, 933 North Cherry Avenue, Tuscon, AZ 85721, USA}
\email{}

\author[0000-0003-3498-2973]{Lee Armus}
\affiliation{IPAC, California Institute of Technology, 1200 E. California Blvd., Pasadena, CA 91125, USA}
\email{}

\author[orcid=0000-0003-0415-0121,gname=Manning,sname=Sinclaire M.]{Sinclaire M. Manning}
\affiliation{Department of Astronomy, University of Massachusetts, Amherst, MA 01003, USA}
\email{}

\author[orcid=0000-0003-3881-1397,gname=Cooper,sname=Olivia]{Olivia Cooper}
\affiliation{Department for Astrophysical \& Planetary Science, University of Colorado, Boulder, CO 80309, USA}
\email{}

\author[orcid=0000-0003-0699-6083, gname=Díaz-Santos,sname=Tanio] {Tanio Díaz-Santos}
\affiliation{Institutes of Computer Science and Astrophysics, Foundation for Research and Technology Hellas (FORTH), 100 Nikolaou Plastira str., Vassilika Vouton, Heraklion, 70013, Greece}
\email{}

\author[0000-0002-4690-4502]{Miriam Eleazer}
\affiliation{Department of Astronomy, University of Massachusetts, Amherst, MA 01003, USA}
\email{}

\author[0000-0002-0930-6466]{Caitlin M. Casey}
\affiliation{Department of Physics, University of California Santa Barbara, Santa Barbara, CA, USA}
\email{}

\author[0000-0001-8490-6632]{Thomas S.-Y. Lai}
\affiliation{IPAC, California Institute of Technology, 1200 E. California Blvd., Pasadena, CA 91125, USA}
\email{}

\author[orcid=0000-0001-8519-1130, gname=Finkelstein,sname=Steven] {Steven L. Finkelstein}
\affiliation{Department of Astronomy, The University of Texas at Austin, 2515 Speedway Boulevard, Austin, TX 78712, USA}
\affiliation{Cosmic Frontier Center, The University of Texas at Austin, Austin, TX 78712}
\email{}

\author[0000-0002-8502-8947]{Leonid Sajkov}
\affiliation{Kavli Institute for Particle Astrophysics and Cosmology, Stanford University, 452 Lomita Mall, Stanford, CA 94305}
\affiliation{Department of Physics, Stanford University, 382 Via Pueblo Mall, Stanford, CA 94305}
\email{}

\author[0000-0002-5537-8110]{Allison Kirkpatrick}
\affiliation{Department of Physics \& Astronomy, University of Kansas, Lawrence, KS 66045, USA}
\email{}

\author[orcid=0000-0003-1282-7454,gname=Taylor,sname=Anthony]{Anthony Taylor}
\affiliation{Department of Astronomy, The University of Texas at Austin, 2515 Speedway Boulevard, Austin, TX 78712, USA}
\affiliation{Cosmic Frontier Center, The University of Texas at Austin, Austin, TX 78712}
\email{}

\begin{abstract}

Near-infrared hydrogen recombination lines can provide robust star-formation rates if the effects of dust attenuation continue to diminish towards longer wavelengths. To quantify near-infrared attenuation and ultimately measure accurate star-formation rates (SFRs) at the height of cosmic star-formation, we present Pa$\alpha$, Br$\alpha$, and Br$\beta$ detections in $27$ luminous infrared galaxies at $z\sim 1-2$, including pure star-forming galaxies as well as dust-obscure Active Galactic Nuclei (AGN). Using combined spectra from the JWST MIRI (Mid-Infrared Instrument) LRS (Low Resolution Spectrometer) and the Spitzer Space Telescope IRS we quantify the star formation rates of our sample using a sub-sample of 13 galaxies detected in Pa$\alpha$ emission ($1.87 \mu$m), 14 galaxies detected in Br$\alpha$ ($4.05 \mu$m), and 12 galaxies detected in Br$\beta$ ($2.63 \mu$m). By combining multiple near-infrared recombination lines in individual galaxies we calculate nebular attenuation, finding $A_\lambda\sim3$ mag at $\sim2\,\mu$m and $A_\lambda\sim1$ mag at $\sim4\,\mu$m. Nebular lines trace the SFR on $\sim 10$ Myr timescales, which we compare to the total infrared luminosity ($L_{\rm IR}$) which probes the star formation over the last $\sim 100$ Myr. SFRs measured from $L_{\rm IR}$ using common conversions are systematically higher than the SFRs derived from attenuation-corrected nebular line measurements by a factor of $\sim2.5$. This could be indicative of a declining or static star formation history with a past burst or systematic effects on $\rm SFR_{\rm IR}$ such as dust heating from evolved stellar populations, and/or an active galactic nucleus.

\end{abstract}


\section{Introduction} \label{Introduction}

\begin{figure}
    \centering
    \includegraphics[width=\linewidth]{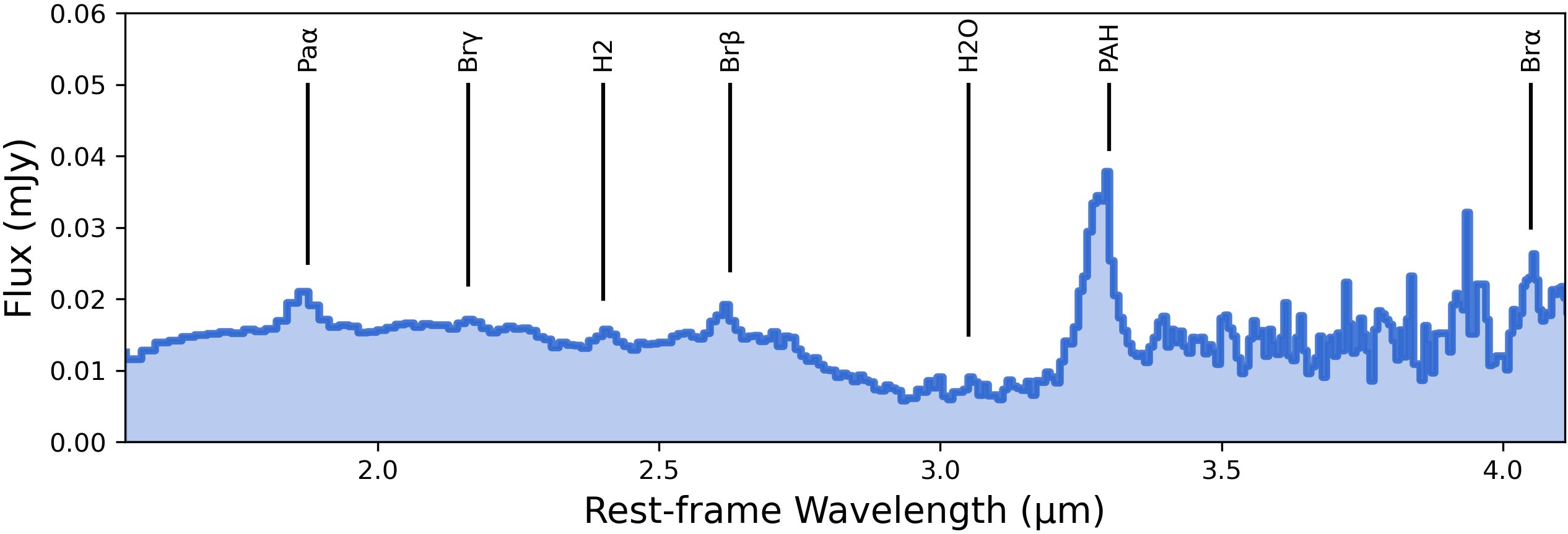}
    \caption{JWST MIRI/LRS spectra of GN-IRS-7 ($\lambda_{\rm obs}=5-14 \mu$m) at $z_{spec}=1.998$, the only galaxy in our sample with the wavelength coverage in JWST to have Pa$\alpha$, Br$\alpha$, and Br$\beta$ emission lines. Due to the noise at Br$\alpha$, it fails our SNR $>3$ cut. We use the observed line fluxes from these hydrogen recombination lines throughout the sample to measure dust attenuation in the individual galaxies.}
    \label{Fig:GN_IRS7}
\end{figure}

Nearly 10 billion years ago ($z\sim 2$) galaxies were forming more stars and growing their supermassive black holes more rapidly than at any other time in the Universe. This is observed as a peak in the star formation rate density (SFRD, \citealt{Madau_2014, Zavala_2021}) and black hole accretion rate density (BHARD, \citealt{Madau_2014, Kim_2024, Porras-Valverde_2026}). During this time, dusty luminous infrared galaxies (LIRGs, $L_{\rm IR} \equiv 10^{11} - 10^{12} L_{\odot}$) and ultra luminous infrared galaxies (ULIRGs, $L_{\rm IR} \equiv 10^{12} - 10^{13} L_{\odot}$) dominate the SFRD \citep{Zavala_2021}.

Star-formation rates (SFRs) in heavily dust-attenuated galaxies are best measured in the mid-infrared where the effects of dust attenuation on hydrogen recombination lines are minimal \citep[e.g.,][]{Alberts_2026} and thermal emission from dust heated by starlight can be observed directly \citep{Kennicutt_1998, Calzetti_2013}, or in the radio from extinction-independent free-free emission \citep{Murphy_2011}. These different star-formation rate indicators have their own benefits and drawbacks, and probe the star-formation rate averaged over different timescales. For instance, hydrogen recombination lines trace \ion{H}{2} regions photoionized by O stars with lifetimes on timescales of $\sim 10$ Myr \citep{Madau_2014}. Dust attenuation corrections are needed for accurate SFRs derived from hydrogen lines, even in the near-infrared, owing to non-negligible opacities \citep{LoFaro_2017, Reddy_2020, Reddy_2025, Wozniak_2026}. The integrated $\lambda_{\rm rest}=8-1000\,\mu$m infrared luminosity ($L_{\rm IR}$) traces the SFR averaged over $\sim100$ Myr and is more robust against attenuation effects, but can suffer from degenerate heating effects by old stellar populations and supermassive black holes \citep{Hayward_2014, McKinney_2021, Bardati2026}. Collecting multiple SFR diagnostics in galaxies is one way to mitigate these systematic uncertainties \citep[e.g.,][]{Murphy_2011}. This has been historically challenging for dust-obscured galaxies due to high rest-frame optical attenuation \citep{Casey2014b}. Thanks to its sensitivity, JWST can now access these features in galaxies with high $V-$band attenuation ($A_V\gtrsim1$, e.g., \citealt{Cooper_2024}), and can reach rest-frame near-infrared recombination lines owing to its wavelength coverage \citep{Neufeld_2024, Alberts_2026}. 

In this work, we use the JWST Mid-Infrared Instrument (MIRI) Low Resolution Spectrometer (LRS) to measure SFRs and near-infrared attenuation in a sample of 27 massive, infrared-luminous galaxies ($L_{\rm IR}/L_{\odot}\sim10^{11}-10^{13}$) at $z=0.6-2.2$ from \cite{McKinney_2025}. This sample includes pure star-forming galaxies,  heavily obscured Active Galactic Nuclei (AGN), and composites. We compare SFRs from Paschen $\alpha$ (Pa$\alpha$), Brackett $\alpha$ (Br$\alpha$), and Brackett $\beta$ (Br$\beta$) to IR-derived SFRs to characterize nebular attenuation and SFRs at different timescales in the most actively star-forming galaxies at cosmic noon ($z\sim1-3$). 

Section \ref{Data} summarizes the sample selection and data reduction from \cite{McKinney_2025}. In Section \ref{Analysis} we describe our analysis of the Pa$\alpha$, Br$\alpha$, and Br$\beta$ emission lines used to measure SFRs and near-IR dust attenuation. In Section \ref{Discussion} we discuss and compare these SFRs to each other, and we summarize our findings in Section \ref{Conclusions}. Throughout this paper we assume a Chabrier Initial Mass Function (IMF) \citep{Chabrier_2003} and adopt a standard cosmology of $\Omega_\Lambda=0.7$, $\Omega_M=0.3$, and $H_0=70$ $\rm{km}~s^{-1}~Mpc^{-1}$. All magnitudes correspond to the AB magnitude system \citep{Oke1974}.

\begin{figure*}
    \centering
    \includegraphics[width=1\linewidth]{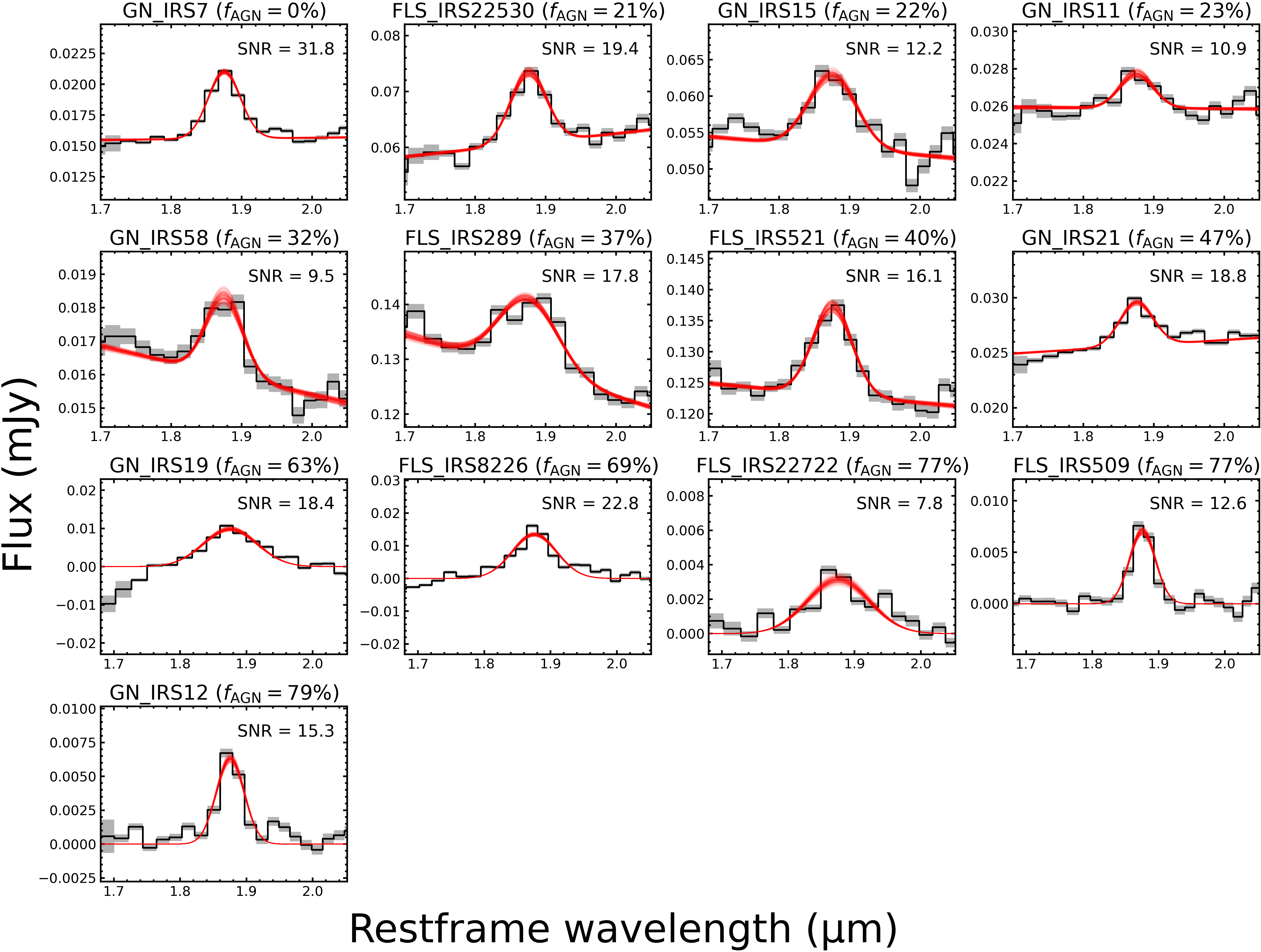}
    \caption{Pa$\alpha$ \texttt{emcee} fits for our subsample of galaxies that have SNR $>3$ Pa$\alpha$ detections, ordered by the fraction of mid-infrared emission attribuated to a dust-obscured AGN ($f_{\rm AGN}$). The observed spectra are in black, and have been continuum-subtracted in sources with $f_{\rm AGN}>60\%$ for clarity due to the steeply-rising near-IR continuum Pa$\alpha$ sits on top of. The \texttt{emcee} modeled emission are the red gaussian lines showing the top $10\%$ of fits. These fits incorporate both the line spread function of JWST MIRI and are resampled to the spectral resolution of LRS. 
    }
    \label{Fig:PaFits}
\end{figure*}

\section{Data} \label{Data}

\subsection{Sample Selection}
In this work we analyze a subset of galaxies with MIRI LRS spectra from the JWST Cycle 2 Program $\#3224$ (PI: McKinney). The JWST data used in this paper can be found in the Mikulski Archive for Space Telescopes (MAST):\dataset[10.17909/18v0-8068]{http://dx.doi.org/10.17909/18v0-8068}. The survey design, strategy, and data reduction are described in \cite{McKinney_2025}. In summary, JWST PID $\#3224$ targeted 37 dusty galaxies with redshifts $0.6 < z < 2.2$ and stellar masses $10^{10}-10^{11.5}M_{\odot}$ using the JWST MIRI Low Resolution Spectrometer (LRS) in combination with archival Spitzer IRS spectra to cover an observed wavelength range of $\lambda_{\rm obs}=5-38 \mu$m. 
These galaxies span a range of AGN contribution from pure dust-obscured star-forming galaxies to highly obscured AGN. The AGN contribution is quantified by $f_{\rm AGN}$, the fraction of light in the mid-IR ($\lambda_{rest}\sim 5-12 \mu$m) attributed to thermal dust emission from a hot dusty torus around the AGN, which is accurate to $10\%$ \citep{Pope_2008, Kirkpatrick_2015}.
This sample was selected from a parent sample of 343 LIRGs described in \cite{Kirkpatrick_2012} and we utilize the $f_{\rm AGN}$ measurements made in \cite{Kirkpatrick_2012, Kirkpatrick_2015, Kirkpatrick_2017}. The JWST data reduction of the sample is discussed in detail in \cite{McKinney_2025}. As a brief summary, the sample is reduced using the JWST pipeline (version 1.15.1, \citealt{Bushouse_2025}) with custom background subtraction in place of the standard Stage 3 routines. 1D spectra are extracted from the background-subtracted 2D spectra using a custom PSF-weighted aperture for each source, and are then normalized to Spitzer IRAC photometry and IRS spectroscopy. 

\begin{figure*}
    \centering
    \includegraphics[width=1\linewidth]{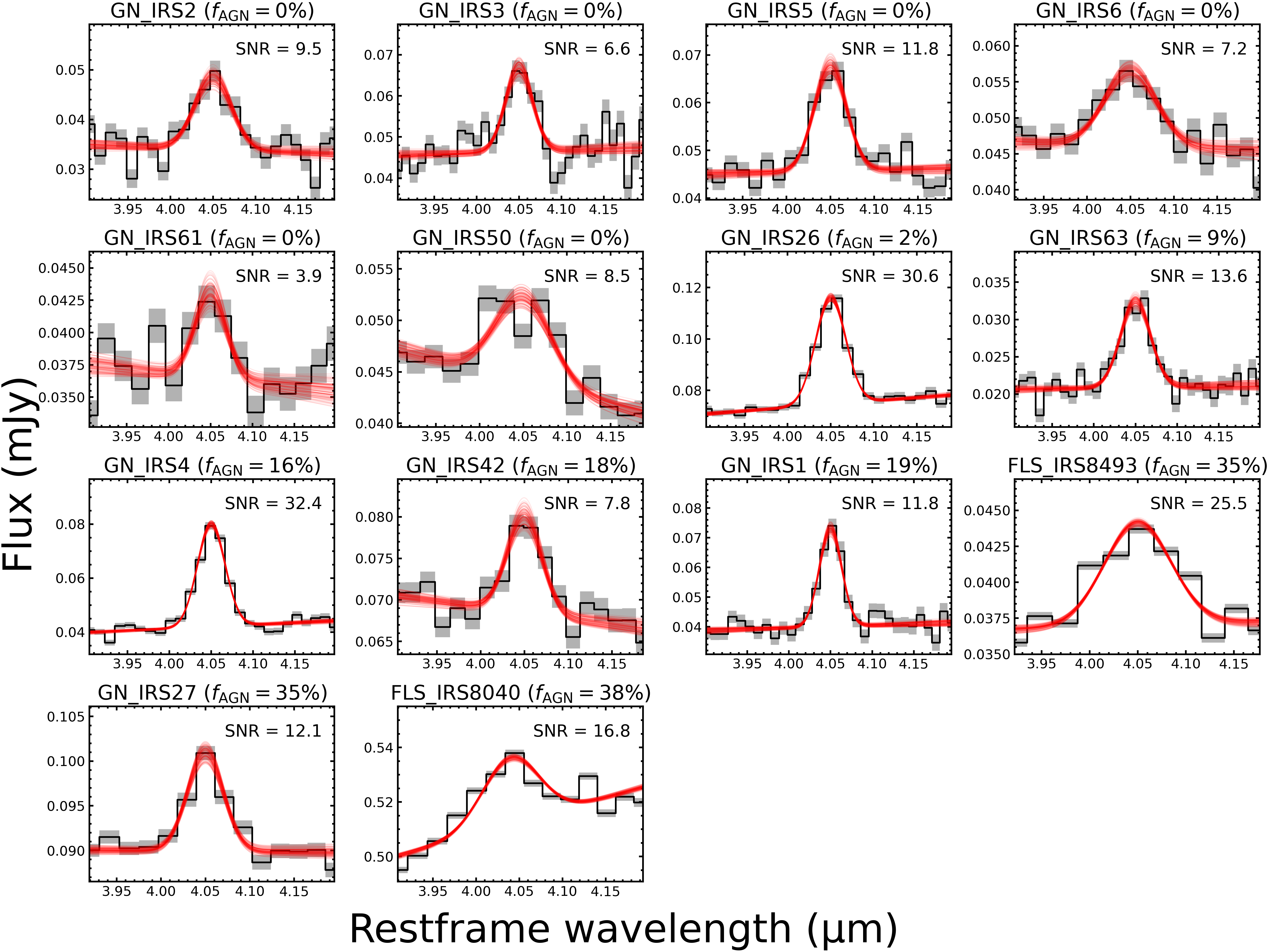}
    \caption{Br$\alpha$ \texttt{emcee} fits for our subsample of galaxies that have a SNR $>3$ Br$\alpha$ detection in order of $f_{\rm AGN}$. The observed spectra are in black. The \texttt{emcee} modeled emission are the red gaussian lines showing the top $10\%$ of fits. These fits incorporate both the line spread function of JWST MIRI and the spectral resolution of LRS.}
    \label{Fig:BrFits}
\end{figure*}

From the parent sample of 37 galaxies from \cite{McKinney_2025} we select all sources with a significant (signal-to-noise ratio, SNR$>3$) Pa$\alpha$ ($1.87 \mu$m), Br$\alpha$ ($4.05\mu$m), or Br$\beta$ ($2.63 \mu$m) emission line. The number of galaxies in each subsample and their redshift, stellar mass, and $f_{\rm AGN}$ are described in Table~\ref{tab:sample_properties}.

\begin{deluxetable}{lcccc}
    \tablecaption{Properties of the Pa$\alpha$, Br$\alpha$, and Br$\beta$ sub-samples. \label{tab:sample_properties}}
    \tablehead{
        \colhead{Sample} & \colhead{Amount} & \colhead{Redshift} & \colhead{$\log(M_*/M_\odot)$} & \colhead{$f_{\rm AGN}$}
    }
    \startdata
    Pa$\alpha$ & $13$ & $1.8-2.2$ & $10.1-11.6$ & $0-79\%$ \\
    Br$\alpha$ & $14$ & $0.6-1.5$ & $10.2-11.5$ & $0-38\%$ \\
    Br$\beta$  & $12$ & $1.2-2.1$ & $10.7-11.6$ & $0-79\%$ \\
    \enddata
\end{deluxetable}

\begin{figure*}
    \centering
    \includegraphics[width=1\linewidth]{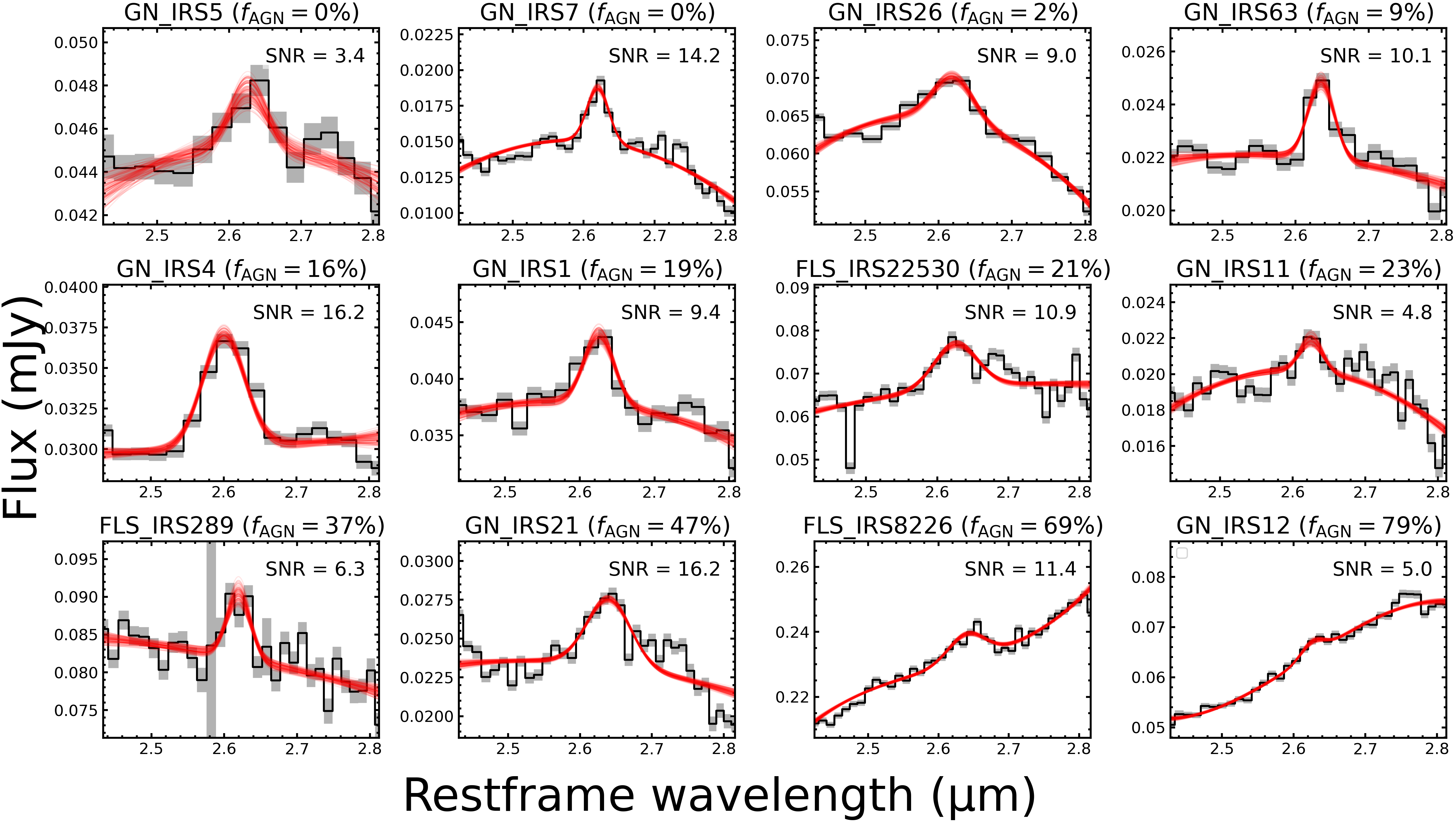}
    \caption{Br$\beta$ \texttt{emcee} fits for our subsample of galaxies that have a SNR $>3$ Br$\beta$ detection in order of $f_{\rm AGN}$. The observed spectra are in black. The \texttt{emcee} modeled emission are the red gaussian lines showing the top $10\%$ of fits. These fits incorporate both the line spread function of JWST MIRI and the spectral resolution of LRS.}
    \label{Fig:BrBFits}
\end{figure*}

Due to the wavelength coverage of MIRI LRS ($\lambda_{\rm obs}=5-14 \mu$m), the Pa$\alpha$ subset consists of the galaxies in the parent sample with the highest redshifts. The sample of Br$\alpha$ emitting galaxies span a relatively smaller redshift range. At higher redshifts, Br$\alpha$ falls in Spitzer IRS and cannot be detected due to the limited sensitivity and resolving power of those data. 
The Pa$\alpha$ and Br$\beta$ sub-samples span a larger range in $f_{\rm AGN}$ than the subset of sources detected in Br$\alpha$. This arises from the original selection criterion of the parent sample tied to Spitzer MIPS 24$\,\mu$m fluxes, which combined with the redshifting of the Pa$\alpha$ line and the blue edge of LRS, preferentially selects the more luminous galaxies at $z>1.65$ which are more likely to harbor a dust-obscured AGN \citep{Kirkpatrick_2015}. Nevertheless, star-formation dominated galaxies with $f_{\rm AGN}<30\%$ are represented in each sub-sample. 
Only one galaxy (GN-IRS-7, shown in Figure~\ref{Fig:GN_IRS7}) sits at a redshift of $z=1.998$ wherein Pa$\alpha$, Br$\alpha$, and Br$\beta$ fall within the LRS wavelength coverage but the Br$\alpha$ emission has a much lower SNR than the other two lines due to the declining sensitivity of LRS beyond $\sim13\,\mu$m.

\section{Analysis} \label{Analysis}

\subsection{Total Stellar Mass}
We fit all sources in the sample using the spectral energy distribution (SED) fitting code \texttt{CIGALE} \citep{Boquien2019}, assuming a double-exponential SFH with $\tau_{\rm main}\in[0.1,3]$ Gyr and $\tau_{\rm burst}\in[-0.3,0.01]$ Gyr. We model an AGN component using the \texttt{SKIRTOR} model library \citep{Stalevski2016}, and the IR SED using the combined modified blackbody and mid-IR power law model from \cite{Casey2012}. These fits make use of near- through far-infrared photometry from $\sim1-500\,\mu$m, including \textit{J-} and \textit{K-}band photometry from VLT/ISAAC \citep{Retzlaff_2010}, Spitzer IRAC $3.6,4.5,5.8,8.0 \mu$m, MIPS $24\mu$m; Herschel PACS $70,100,160\mu$m and SPIRE $250, 350, 500\mu$m \citep{Kirkpatrick_2015}. We find no systematic offset between the stellar masses estimated this way and those from prior works on the same galaxies \citep{Kirkpatrick_2017}. 

\subsection{Emission Line Modeling\label{sec:linemodels}}
To measure SFRs from the hydrogen recombination lines, we fit Pa$\alpha$, Br$\alpha$, and Br$\beta$ using \texttt{emcee} \citep{emcee}, an affine invariant Markov Chain Monte Carlo ensemble sampler. We fix the redshift and line center of each source based on the redshifts measured in \cite{McKinney_2025} from the Pa$\alpha$ and Br$\alpha$ emission lines. We model the lines as a 1D Gaussian with a linear or polynomial continuum, dependent on the effect of AGN on the continuum shape. We adopt a 2nd order polynomial continuum for sources with $f_{\rm AGN}>65\%$. We restrict to fitting just the local continuum within $\Delta\lambda\approx0.15\,\mu$m which we found to best capture the continuum behavior with minimal contamination from nearby features. In addition, we account for the effects of the MIRI LRS line spread function by convolving model spectra with a Gaussian kernel where the width is determined from the P750L spectral resolving power using the wavelength-dependent dispersion and resolution, as obtained from the JWST Pandeia reference data system (TRDS) v5.1.
Finally, models are re-gridded to the LRS wavelength resolution using a flux-conserving resampling algorithm implemented in \texttt{specutils} \citep{specutils_fluxconservingresampler}. We note that this fitting method works particularly well for Pa$\alpha$ and Br$\alpha$ due to the well-behaved local linear continuum. The continuum around Br$\beta$ is much more complicated. Br$\beta$ is between two absorption features (as can be seen in Figure~\ref{Fig:GN_IRS7}) caused by stellar atmospheres and dense gas and dust making the continuum non-linear in many instances (see Figure~\ref{Fig:BrBFits}). We fit the continuum around Br$\beta$ with a 2nd order polynomial which we found in testing to preform the best continuum characterization upon visual inspection. For sources with $f_{\rm AGN}>60\%$ we adopt a 3rd order polynomial to account for the absorption features and the steep continuum caused by the AGN. The continuum is well-separated by Br$\beta$ in the majority of our spectra, and we note that the precise continuum model does not significantly impact the inferred line luminosities. 
Note that for sources dominated by AGN ($f_{\rm AGN}>60\%$), the emission lines do not appear as prominent as in their star-formation dominated counterparts. This is due to a combination of the aggressive smoothing from the LSF and the steeply-rising power-law nature of the continuum. For visual purposes, we have continuum-subtracted the AGN dominated sources in Fig.~\ref{Fig:PaFits}. Line uncertainties and the subsequent SNR ratios shown on Figures~\ref{Fig:PaFits} and \ref{Fig:BrFits} are taken from the empirical spectral noise, which is the dominant error term for Pa$\alpha$ and Br$\alpha$ due to the well-behaved continuum. Br$\beta$ lines fluxes are dominated by the model uncertainties, which the reported SNRs incorporate.

LRS has a spectral resolution of $R\approx100$, which means that the emission lines in our targets are not spectroscopically resolved. As a result, we do not have constraint on velocity widths. We therefore restrict line widths to the narrow range of $300\le v\le 500\,{\rm km\,s^{-1}}$ as is typical for the observed rotational velocities of galaxies with similar total stellar mass and IR luminosity to our targets \citeg{Birkin_2024, Cooper_2024}. 
The best fit models to the Pa$\alpha$, Br$\alpha$, and Br$\beta$ emission lines can be seen in Figure~\ref{Fig:PaFits},  Figure~\ref{Fig:BrFits}, and Figure~\ref{Fig:BrBFits} respectively with each emission line's corresponding SNR. 

\begin{figure}
    \centering
    \includegraphics[width=1\linewidth]{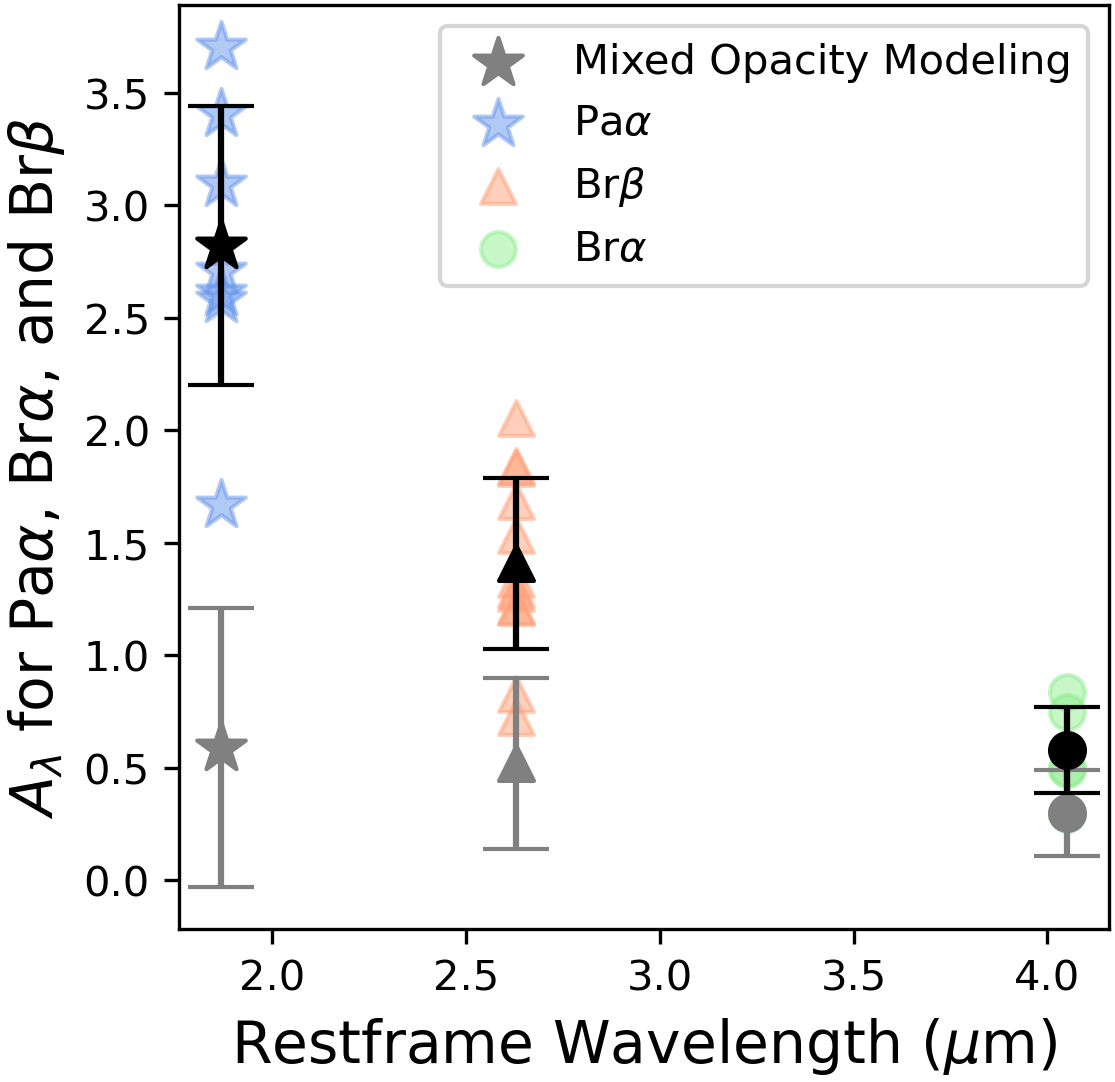}
    \caption{Measured attenuation from $1.8-4.05\mu$m across our sample from galaxies detected in two ore more hydrogen transitions. Pa$\alpha$ is denoted by stars, Br$\beta$ by triangles, and Br$\alpha$ by circles. The average of the mixed opacity modeling from \cite{McKinney_2025} is shown in gray which represents the attenuation correction applied to the stellar and dust continuum at these wavelengths. Each individually measured nebular line attenuation is plotted in color with the average of each sub-sample in black. The difference between the continuum and nebular attenuation is $\sim 2$ magnitudes at Pa$\alpha$, and consistent with one another by Br$\alpha$ at $4.05\mu$m.}
    \label{Fig:AttVsWave}
\end{figure} 

\subsection{Dust Attenuation\label{sec:dustAttenuation}}
To correct for the effects of dust attenuation on emission lines measured from the JWST spectra, we begin by considering the opacity curves described in \cite{McKinney_2025} which were obtained by fitting the LRS spectra with wavelengths of $3-12 \mu$m and photometric constraints out to rest-frame $24 \mu$m. These opacity curves model all of the dust in the systems as a mixed dust geometry wherein the dust is mixed homogeneously with the emitting medium. The dust extinction profile comes from \cite{Weingartner_2001} which assumes Milky Way-like dust \citep{Cardelli_1989, Clarke_2025}, that is then scaled to the opacity of the $9.7 \mu$m silicate absorption feature of each source to infer the shape of the extinction profile down to short wavelengths. While this is appropriate for deriving dust attenuation corrections to the the Polycyclic Aromatic Hydrocarbon features \citeg{Lai_2020}, which was the focus of \cite{McKinney_2025}, this assumption likely underestimates the attenuation corrections for nebular emission lines tracing young star-forming regions deeply embedded in dust-rich birth clouds \citep{Calzetti_2000, Charlot_2000}. In addition, the extinction curves from the \cite{McKinney_2025} fits in the rest-frame near-IR rely on extrapolations to bluer wavelengths, further adding to their uncertainty.

Hydrogen recombination line ratios can be used to derive a nebular attenuation correction by comparing to the expected ratio for Case B recombination \citeg{Reddy_2020, Cleri_2022, Prescott_2022, Reddy_2023, Cooper_2024, Sanders_2025, Reddy_2025}. Naturally, this method requires more than one hydrogen recombination line detection.
Our sample includes 7 galaxies with both Pa$\alpha$ and Br$\beta$, and 5 galaxies with both Br$\alpha$ and Br$\beta$. We use these to estimate the attenuation impacting the nebular emission lines by first assuming Case B intensity ratios from \cite{Osterbrock_2006} for $T=10^4K$. We then follow \cite{Calzetti_2000, Wang_2019, Reddy_2020, Reddy_2025}. To estimate the $V-$band attenuation we assume an attenuation law of $A_\lambda \propto \lambda^{-\alpha}$ and $\alpha=2.07\pm0.03$ \citep{Wang_2019}. Our final nebular reddening corrections are insensitive to the choice of $\alpha$.
We measure a mean ($\pm1\sigma$) attenuation of $A_{\rm Pa\alpha}=2.82 \pm 0.61$ mag, $A_{\rm Br\beta}=1.4 \pm 0.39$ mag, and $A_{\rm Br\alpha}=0.58\pm 0.19$ mag. As expected, we measure a decrease in the mean attenuation for progressively longer wavelength emission lines (as can be seen in Figure~\ref{Fig:AttVsWave}). The attenuation at $\lambda_{\rm rest}=4.05\,\mu$m remains non-zero with $\sim0.6$ mag, similar of the near-infrared attenuation found for local ULIRGs \citep{Murphy_2001,Dannerbauer2005,Armus2007}. Given the extreme obscuration of our sample, even these nebular attenuation measurements may be underestimating the attenuation toward the most deeply embedded star-forming regions in our sources.

\begin{figure}
    \centering
    \includegraphics[width=1\linewidth]{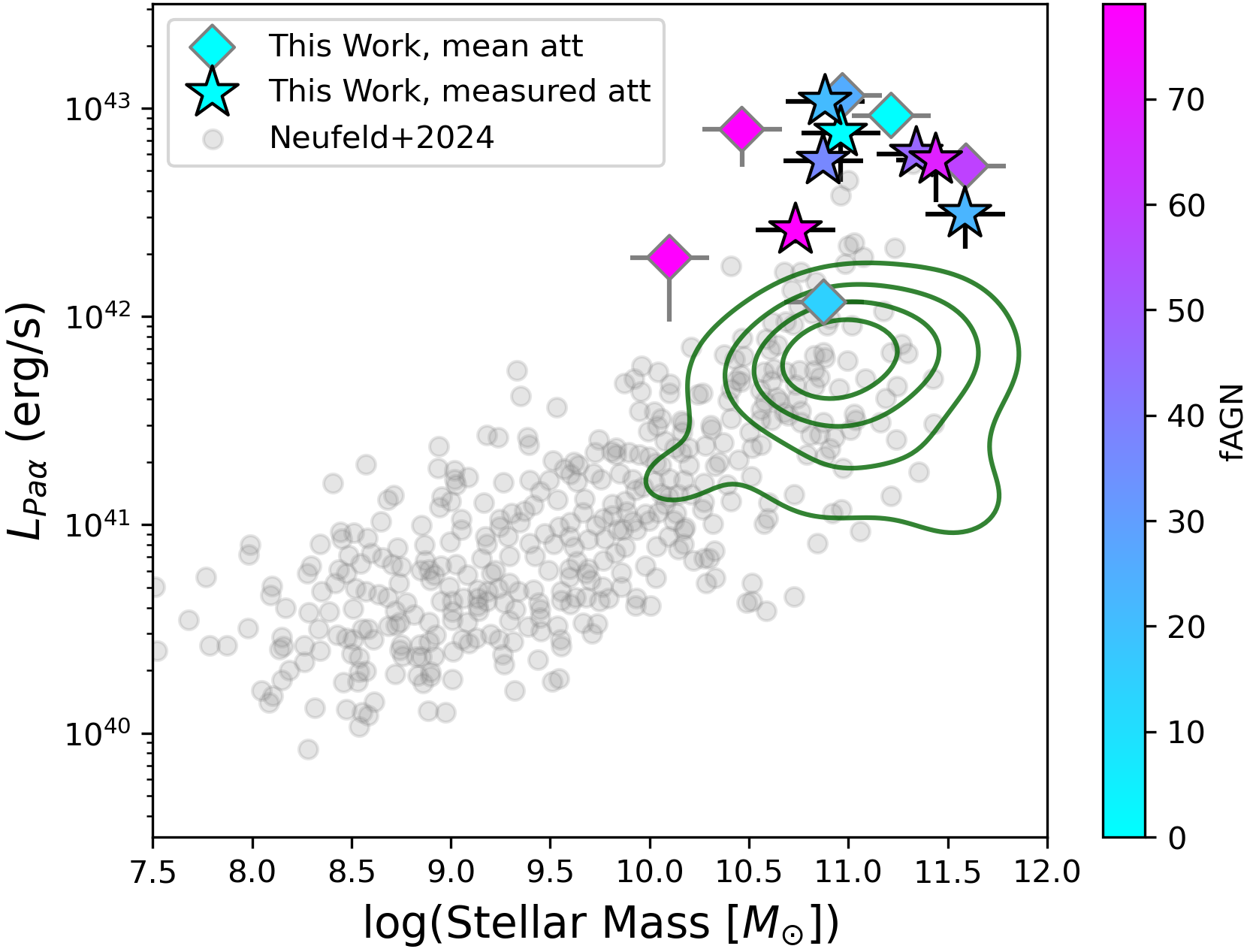}
    \caption{Pa$\alpha$ luminosity and stellar mass for galaxies in our sample compared to that of \cite{Neufeld_2024} (gray circles) which also uses Pa$\alpha$ at $z=1-1.7$ galaxies to measure a SFR. Our galaxies are at the higher mass end of the \cite{Neufeld_2024} sample and are initially consistent with the higher luminosity sources. After dust correction from our nebular derived attenuation (Section~\ref{sec:dustAttenuation}, where the gray outline points utilize the average attenuation of the sources with multiple hydrogen recombination lines), we find that our sources are within the envelope of the sources from \cite{Neufeld_2024}, but that they reach the high luminosity end of the sample. The green contours show the luminosity distribution of sources without dust correction, indicating that the nebular dust attenuation correction systematically increased Pa$\alpha$ luminosities.}
    \label{Fig:NeufeldComp}
\end{figure}

From these same galaxies, using the mixed opacity modeling from \cite{McKinney_2025}, the mean ($\pm1\sigma$) attenuation is $A_{\rm Pa\alpha}=0.59 \pm 0.55$, $A_{\rm Br\beta}=0.52 \pm 0.39$, and $A_{\rm Br\alpha}=0.30\pm 0.17$. These attenuation values are significantly lower than what we calculate from the nebular lines, by up to a factor of $\sim4.5$ at Pa$\alpha$ and $\sim2$ at Br$\alpha$. This is seen in Figure~\ref{Fig:AttVsWave} which shows the average attenuation from the mixed opacity modeling compared to the measured attenuation from the emission lines and their corresponding average. Pa$\alpha$ is the most affected by the new attenuation measurements, and Br$\alpha$ is the least sensitive, consistent with a steeply rising attenuation curve.
Overall this trend naturally arises from differential reddening, where the nebular emission from young, obscured \ion{H}{2} regions experiences more attenuation than stars mixed in the diffuse ISM. 
The relative offset between these two attenuation corrections yields an average differential reddening of ${\rm E(B-V)_{gas}-E(B-V)_{ISM}}= 0.70\pm0.20, 0.27\pm 0.12, 0.09\pm 0.06$ 
for Pa$\alpha$, Br$\beta$, and Br$\alpha$ respectively, assuming $R_V=3.1$ \citep{Calzetti_2000}. This is comparable, particularly with Br$\alpha$ and Br$\beta$, to what is found for the differential reddening between nebular lines and stellar continuum for massive ($M_*>10^{10}\,M_\odot$) $z=2.7-4$ galaxies with NIRSpec detections of H$\alpha$, H$\beta$, and H$\gamma$ \citep{Woodrum_2025,Karthikeyan2026}. Thus we find that the effects of differential reddening remain significant up to Br$\alpha$ in the most massive, dust-obscured galaxies at cosmic noon, as has been observed in their local counterparts \citep{Dannerbauer2005,Armus2007}. 

\subsection{Star Formation Rates from Pa$\alpha$, Br$\alpha$, and Br$\beta$ \label{sec:sfrs}}
\begin{figure}
    \centering
    \includegraphics[width=1\linewidth]{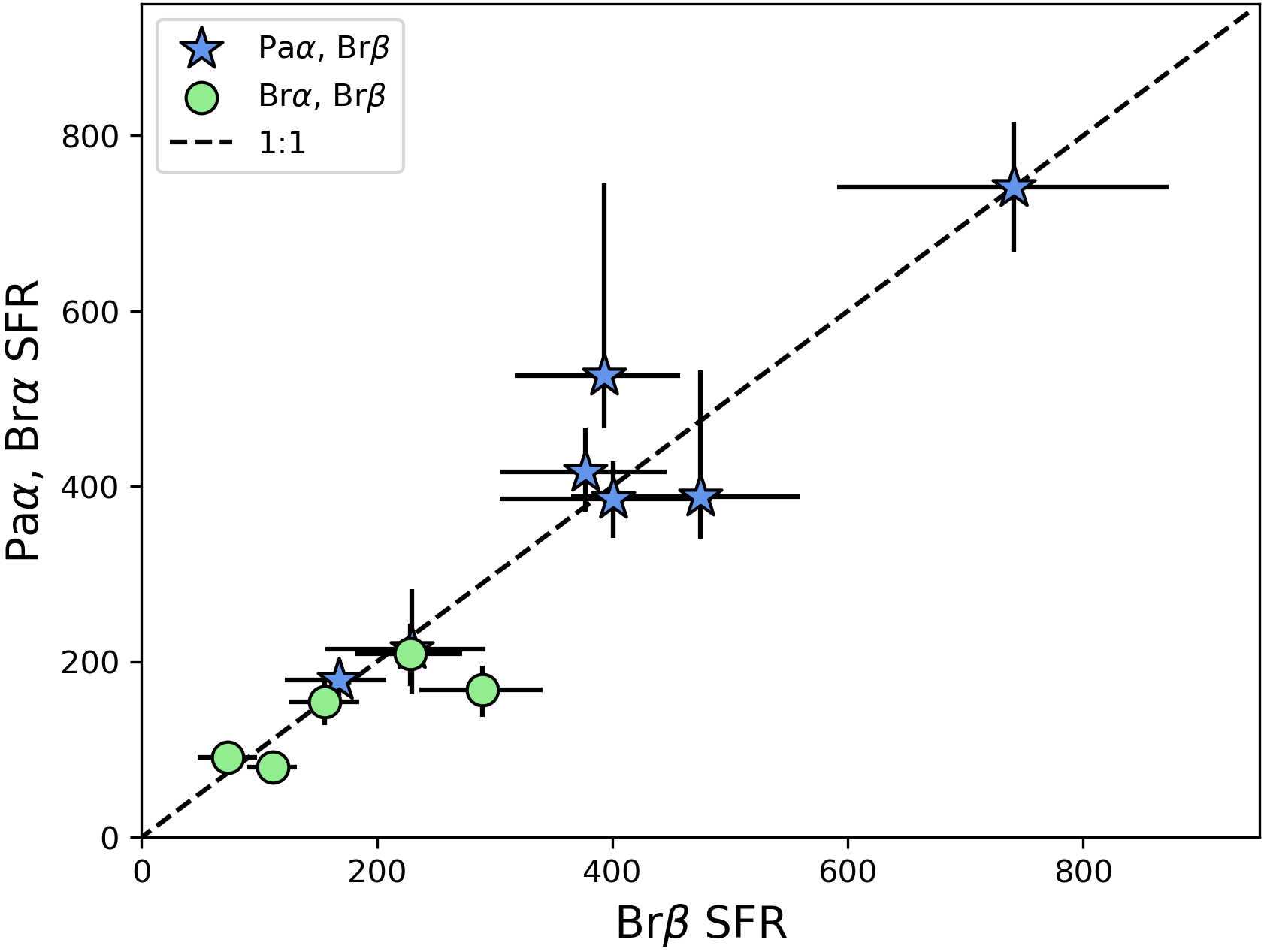}
    \caption{SFR comparison for the galaxies that have multiple hydrogen recombination lines. Galaxies that have both Pa$\alpha$ and Br$\beta$ are denoted by blue stars while sources with Br$\alpha$ and Br$\beta$ are denoted by green circles. There is a very strong correlation between the SFRs measured using Pa$\alpha$, Br$\alpha$ and Br$\beta$ which confirms the use of our attenuation models creating consistency between the SFR indicators.}
    \label{Fig:lineSFRcomp}
\end{figure}

We next use the dust-corrected recombination line luminosities to measure star-formation rates. If the galaxy has multiple hydrogen recombination lines, we use the attenuation measured directly for that source as described in Section~\ref{sec:dustAttenuation}. If the target only has one emission line, we use the average attenuation for the corresponding line measured from the rest of the sample (black points in Figure~\ref{Fig:AttVsWave}). Sources where the nebular attenuation is assumed rather than measured directly are denoted in all figures by a marker with a gray outline (as compared to a black outline). 
We derive SFRs for Pa$\alpha$, Br$\alpha$, and Br$\beta$ using the calibration of the H$\alpha$ SFR in \cite{Kennicutt_2012}. We then use the intrinsic ratios between H$\alpha$ and Pa$\alpha$, Br$\alpha$, and Br$\beta$ in \cite{Osterbrock_2006}, assuming Case B recombination and $T=10^4$ K to convert the H$\alpha$ SFR formulation to Pa$\alpha$, Br$\alpha$, and Br$\beta$. Because we assume a Chabrier IMF, we convert the Kroupa IMF H$\alpha$ SFR in \cite{Kennicutt_2012} through the IMF relation described by \cite{Stanway_2019}. 
Our final SFR relations are: 
\\
\begin{align}
    {\rm SFR}_{\rm Pa\alpha}[M_{\odot}/{\rm yr}]=6.9\times10^{-41}\,L_{\rm Pa\alpha}\,[{\rm erg\,s^{-1}}]
\end{align}
\begin{align}
    {\rm SFR}_{\rm Br\alpha}[M_{\odot}/{\rm yr}]=2.9\times10^{-40}\,L_{\rm Br\alpha}\,[{\rm erg\,s^{-1}}]
\end{align}
\begin{align}
    {\rm SFR}_{\rm Br\beta}[M_{\odot}/{\rm yr}]=5.2\times10^{-40}\,L_{\rm Br\beta}\,[{\rm erg\,s^{-1}}]
\end{align}
\\
We report the resulting SFRs in Table~\ref{tab:recombination}. As shown in Figure~\ref{Fig:lineSFRcomp} for sources with multiple hydrogen recombination lines, the measured line SFRs are consistent with each other within the uncertainties for SFRs between $10-800 M_{\odot}\rm yr^{-1}$.

We note that we do not apply an AGN-ionization correction to the recombination line SFRs \citep[e.g.,][]{Stone_2022}. While strong nebular line emission can arise from the narrow line region around the AGN, this effect is mitigated by increasing obscuration \citep[e.g.,][]{Hickox_2018}. Indeed, we do not find a trend between the recombination line SFRs and mid-IR AGN fraction, suggesting that the obscured AGN in our sample might not systematically bias the corresponding SFRs high. As a final note, only 5/27 ($19\%$) galaxies in our sample have $f_{\rm AGN}>50\%$, suggesting that this effect (if present) would not bias interpretations of the sample as a whole. 

Due to the longer rest-frame wavelength and relatively fainter line flux of Br$\alpha$, literature measurements of Br$\alpha$ are historically limited to nearby galaxies where the line is accessible with ground-based IR facilities or AKARI \citeg{Depoy1987, Beck_1989, Inami_2018}. Here we present the first compilation of Br$\alpha$ SFRs for $z\sim 1-2$ massive, dust-obscured star-forming and AGN dominated galaxies. The $\rm SFR_{\rm Br\alpha}$ are shown in Table~\ref{tab:recombination} and range from $10-150\,M_{\odot}\,{\rm yr^{-1}}$. As the Pa$\alpha$ and Br$\beta$ sample encompass populations with a higher average $L_{\rm IR}$ than Br$\alpha$, their SFR ranges reach higher values. The $\rm SFR_{\rm Pa\alpha}$ and $\rm SFR_{\rm Br\beta}$ are also shown in Table~\ref{tab:recombination} and range from $80-800\,M_{\odot}\,{\rm yr^{-1}}$ and $70-750\,M_{\odot}\,{\rm yr^{-1}}$ respectively.

Figure~\ref{Fig:NeufeldComp} compares the Pa$\alpha$ luminosities we measure to the sample of \cite{Neufeld_2024} who use NIRspec to detect Pa$\alpha$ at $1.0<z<1.7$ as part of the FRESCO survey \citep{Oesch2023}. Without applying any dust corrections, our sample falls along the massive end of the distribution reported in \cite{Neufeld_2024}. Our fiducial dust-corrected Pa$\alpha$ luminosities (using measured nebular attenuation factors) are on-average larger for fixed stellar mass than the most massive galaxies in \citep{Neufeld_2024}. This is expected as our galaxies were selected as a representative sample of some of the brightest dust obscured sources at $z\sim1-2$ that are also relatively rarer and correspondingly less likely to be found in statistical quantities within $<100$ arcmin$^2$ surveys like FRESCO.

\begin{figure}
    \centering
    \includegraphics[width=1\linewidth]{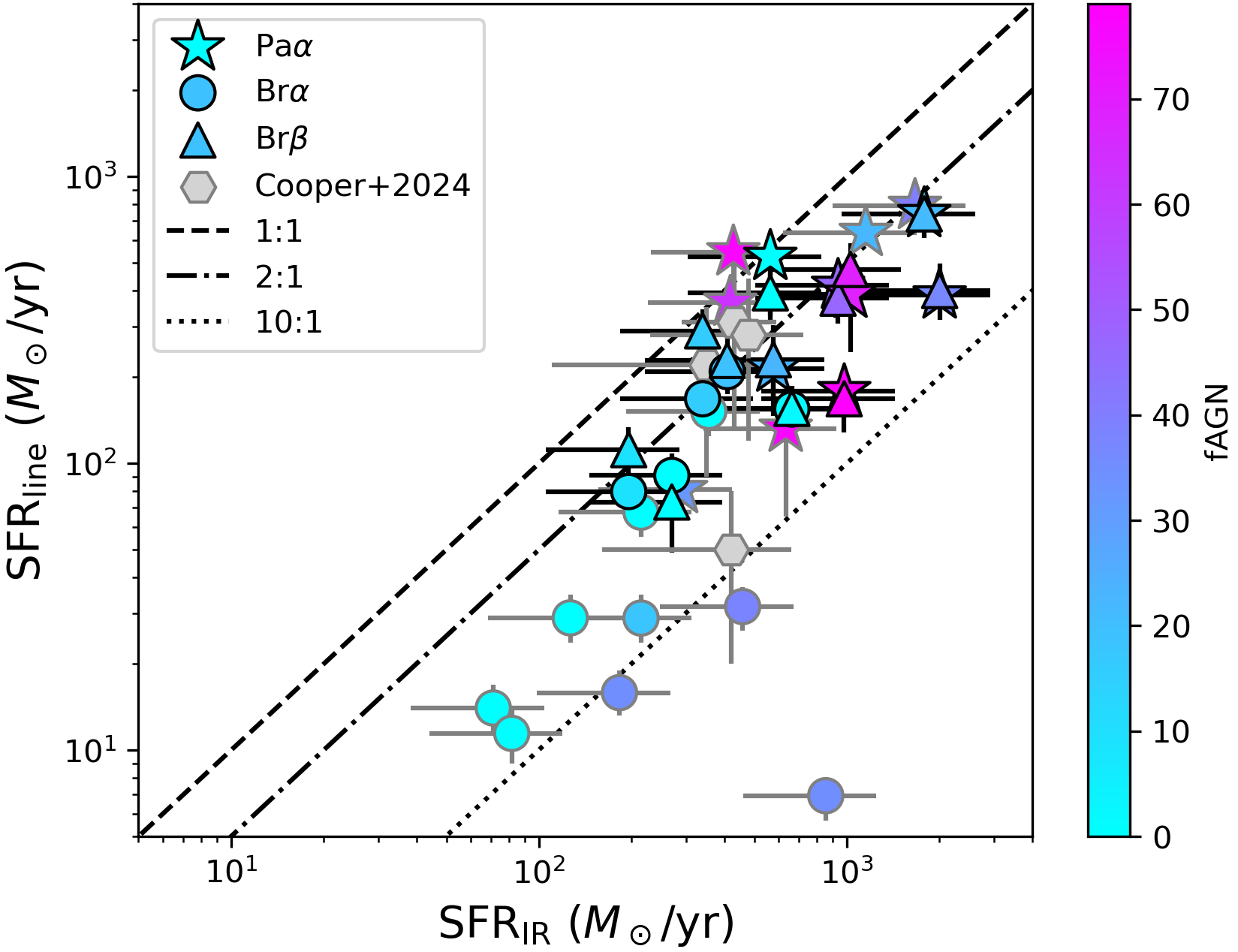}
    \caption{A comparison of the Pa$\alpha$ (stars), Br$\alpha$ (circles), Br$\beta$ (triangles) SFRs ($\sim 10$ Myr) and the $L_{\rm IR}$ SFRs ($\sim 100$ Myr). Sources that had multiple hydrogen recombination lines to measure an individual attenuation have black outlines while the sources that used the mean attenuation for their dust correction have gray outlines. Included are sources from \cite{Cooper_2024} with a similar mass and $L_{\rm IR}$ at $z\sim 2.3-2.7$ which find a similar scatter of SFRs using H$\alpha$ and the $L_{\rm IR}$. The sample falls between zero and one order of magnitude discrepancy of SFR tracers.}
    \label{Fig:BrA_vs_PaA}
\end{figure}

\subsection{Star Formation Rate from the far-IR \label{sec:LIR SFR}}
Using the $L_{\rm IR}$ integrated between $8-1000\mu$m measured from Herschel Space Observatory observations in \cite{Kirkpatrick_2015} we measure SFRs over $\sim 100$ Myr. The SFRs derived from $L_{\rm IR}$ (hereafter SFR$_{\rm IR}$) were measured following \cite{Murphy_2011} as described in the survey overview paper \citep{McKinney_2025}.

\section{Results and Discussion} \label{Discussion}

Using the opacity models derived strictly from the mid-IR continuum modeling \citep{McKinney_2025}, we found that the nebular line derived SFRs were one to two orders of magnitude below $\rm SFR_{IR}$, which we interpret as a significant initial underestimation of the nebular attenuation motivating the analysis described in Section \ref{sec:dustAttenuation}. Using the attenuation derived from multiple hydrogen recombination lines for the sources that have multiple, and the average for the sources that only have one, we compare the nebular line SFRs to the $\rm SFR_{IR}$ for each source in Figure~\ref{Fig:BrA_vs_PaA}. We continue to see an offset between the two different star formation rate indicators, but now the difference is between zero and one order of magnitude with $25\%$ having $\rm SFR_{line}$ that agree with $\rm SFR_{IR}$ within the errors of both measurements.

\begin{figure}
    \centering
    \includegraphics[width=1\linewidth]{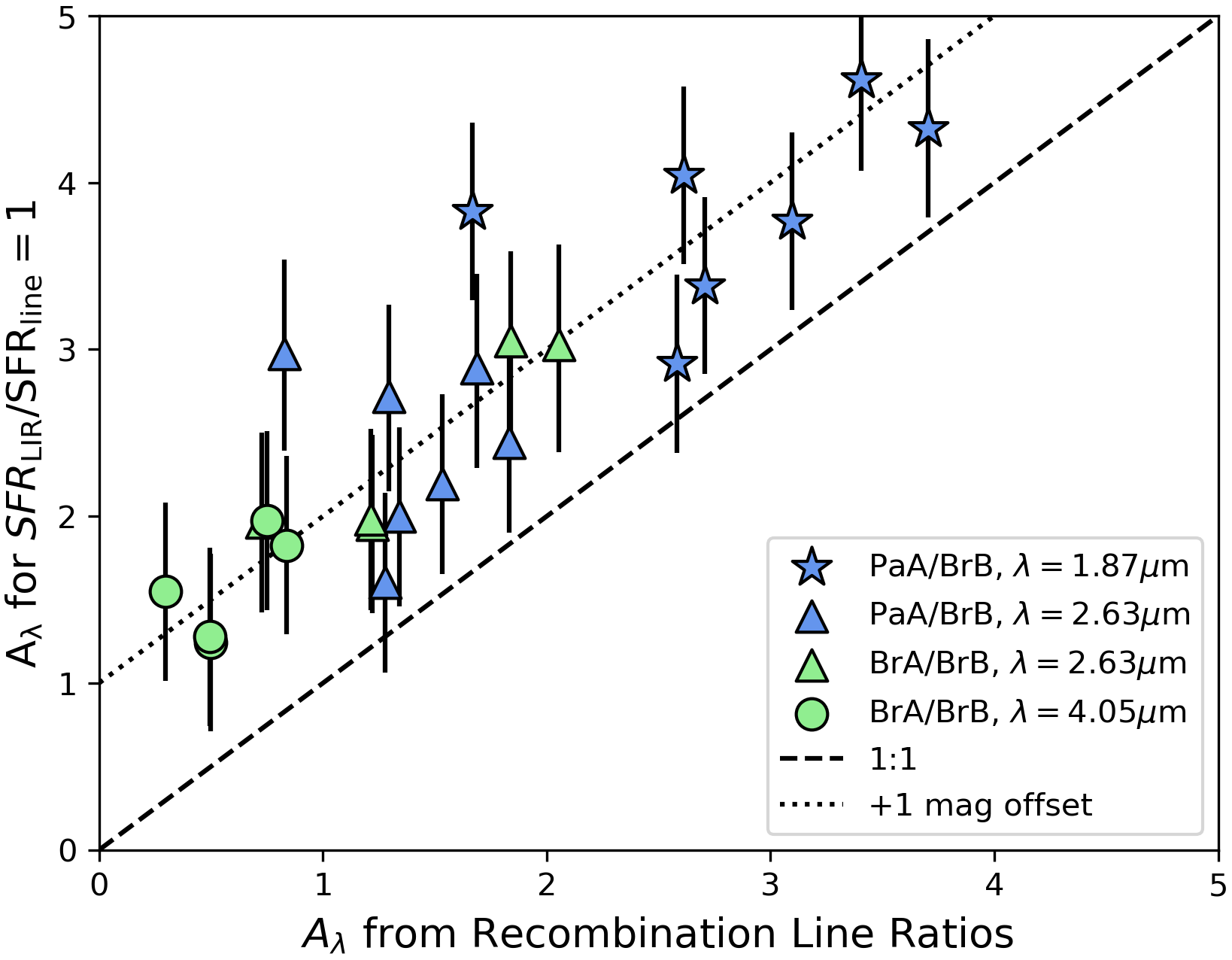}
    \caption{Nebular attenuation derived from the hydrogen recombination lines compared to the amount of attenuation needed to increase the SFRs derived from the emission lines to be the same as the SFRs from the $L_{\rm IR}$ in terms of an attenuation correction factor. The shapes correlate to the SFR used on the y-axis while the colors correlate to the two lines used in the attenuation derivation. We find a systematic offset compared to the 1 mag offset of the y-axis in orange, of our sample with only two sources having an attenuation that puts the recombination SFRs at the same value of $SFR_{\rm IR}$.}
    \label{Fig:AttComp}
\end{figure}

Figure \ref{Fig:AttComp} compares the monochromatic attenuation inferred from nebular line ratios to the attenuation needed for the observed line flux to reproduce $\rm SFR_{\rm IR}$. We find that the attenuation needed for $\rm SFR_{\rm IR}$ and $\rm SFR_{\rm line}$ to agree is systematically greater than what we measure directly by a factor of $\sim1$ mag. This holds for each line pair, meaning the offset does not disappear when using a more transparent part of the spectrum (i.e., Br$\alpha$ vs.\,Pa$\alpha$). \cite{Casey_2026} finds that the $A_V$ inferred from optical/UV continuum underestimates $L_{\rm IR}$ by a factor of $\sim 3$ across most mass ranges out to high redshift. This is consistent with our findings assuming that the attenuated UV/optical continuum \textit{and} nebular lines from stars do not fully capture on-going star-formation in heavily obscured regions.
As discussed in Section \ref{sec:dustAttenuation}, we find comparable differential reddening between nebular lines and stellar continuum to other spectroscopic studies with well-sampled Balmer series \citep{Woodrum_2025,Karthikeyan2026}. 
Therefore, we also consider interpretations including a poorly calibrated and/or contaminated $\rm SFR_{\rm IR}$ measurement, geometric effects, and very dense \ion{H}{2} regions. 

As noted in \cite{Neufeld_2024} using NIRCam/grism measurements, Pa$\alpha$ can exhibit clumpy morphologies. If the $0.51^{\prime\prime}\times4.7^{\prime\prime}$ LRS slit does not capture the full extent of a galaxy's nebular line emission, then we could be missing some amount of Pa$\alpha$, Br$\alpha$, and Br$\beta$ flux even after the aperture corrections which were anchored to unresolved Spitzer/IRAC and IRS peak-up photometry \citep{McKinney_2025}.  We find that there is no correlation between the ratio of $\rm SFR_{Pa\alpha, Br\alpha, Br\beta}$ and $\rm SFR_{IR}$ and the slit-loss correction factors derived in \cite{McKinney_2025} with a Pearson coefficient of $0.034$ and p-value of $0.84$. It is unlikely that slit losses are contributing to the difference in SFRs; however, we acknowledge that they are presently anchored to the continuum which can exhibit different morphological structure than emission lines maps \citep{Neufeld_2024}.

Geometric effects are also discussed in \cite{Cooper_2024} who use NIRSpec MSA slitlets to measure H$\alpha$ across their sample of galaxies of higher redshift than those in this work, but of comparable mass, SFR, and IR luminosity. They note the cold dust emission tracing $L_{\rm IR}$ localized by ALMA is always centered in the MSA slit, indicating that the offset between star-formation rate indicators is likely not driven by missing far-infrared flux outside of the slit. Additional resolved imaging could directly constrain the role of star-formation geometry and clumpy morphologies in our sample. Changes in dust-to-star geometries naturally allow for different emergent attenuation laws \citep{Granato_2000, Calzetti_2000}.

\begin{figure*}
    \centering
    \includegraphics[width=1\linewidth]{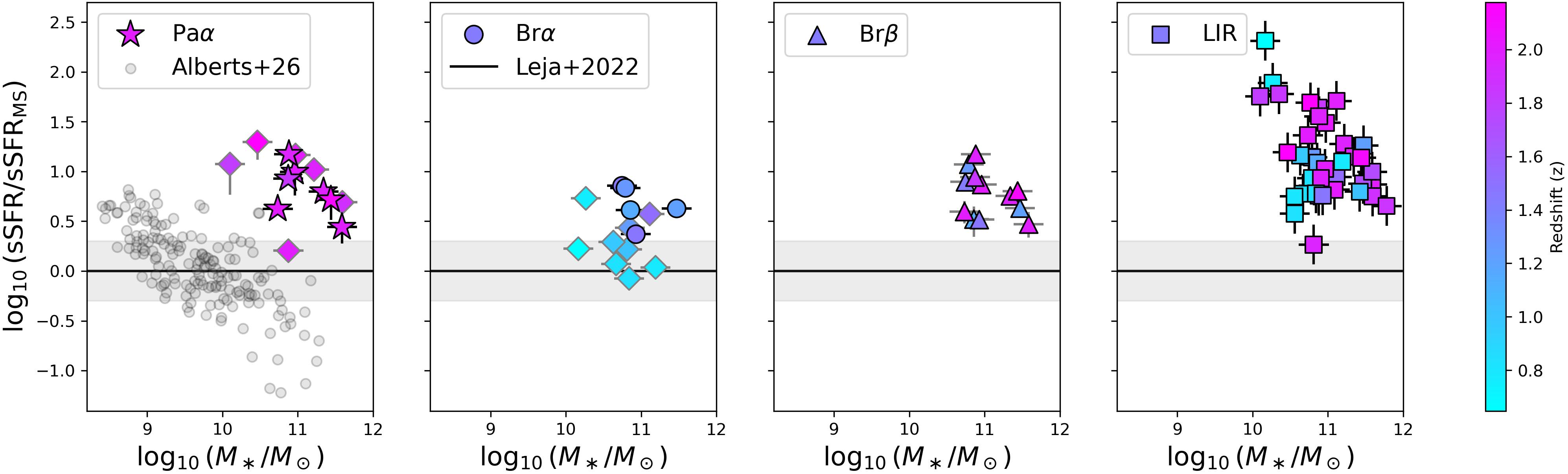}
    \caption{The ratio of each source's specific star formation rate ($\rm sSFR=SFR/M_\odot$) to the main-sequence from \cite{Leja_2022} (solid line with $\pm1\sigma$ scatter) vs. the stellar mass of each source. These star-formation rates are inferred from the dust corrected Pa$\alpha$ (\textit{Left} panel), Br$\alpha$ (\textit{Second} panel), Br$\beta$ (\textit{Third} panel), and $L_{\rm IR}$ (\textit{Right} panel). Sources that had multiple hydrogen recombination lines to measure an individual attenuation have black outlines while the sources that used the mean attenuation for their dust correction are diamonds and have gray outlines. Star-formation rates from nebular lines follow closer to main sequence galaxies of similar mass and redshift, whereas $L_{\rm IR}$ SFR places our sources exclusively above the main-sequence as starburst galaxies. }
    \label{Fig:SF_MS}
\end{figure*}

It is also possible that $\rm SFR_{\rm IR}$ is over-estimated. 
From a theoretical perspective, AGN heating and older stellar populations can bias $L_{\rm IR}$ SFRs high by providing energy from a source unassociated with the UV output of $\sim10-100$ Myr-averaged star formation. 
As shown in \cite{McKinney_2021} and \cite{Bardati2026} using post-processed radiative transfer simulations, dust-enshrouded AGN can power significant cold dust emission such that $L_{\rm IR}$ can be increased by an order of magnitude by reprocessed AGN photons alone. \cite{Hayward_2014} use similar simulation methods to show that $L_{\rm IR}$ can also be boosted by dust heating from evolved stars having formed $>100$ Myr before a burst. Indeed, the largest disagreement between $\rm SFR_{IR}$ and $\rm SFR_{H\alpha}$ observed in the star-forming sample of \cite{Cooper_2024} is associated with the strongest Balmer break and oldest stellar age. Our well-sampled mid-IR spectroscopy mitigates the effects of AGN heating of dust by providing a handle on subtracting out the AGN component of far-IR emission \citeg{Pope_2008,Kirkpatrick_2015}. Dust heating from old stars requires spectroscopy of features like the Balmer break to separate fully which future NIRSpec/PRISM observations may be able to provide. 
That being said, we see the greatest difference between $\rm SFR_{IR}$ and $\rm SFR_{line}$ among the lowest redshift subset ($z<1$) with $f_{\rm AGN}<30\%$ (Fig.\,\ref{Fig:BrA_vs_PaA}), where dust heating by older stars might be the most pronounced in the most evolved galaxies of our sample with the least amount of AGN contamination. However, the contribution of dust heating from evolved stellar populations cannot be quantitatively constrained with the present data.

\cite{Alberts_2026} calibrate photometric mid-IR SFR indicators using Pa$\alpha$ with the FRESCO catalog as used in \cite{Neufeld_2024}, and discusses differences between $\rm SFR_{\rm IR}$ and $\rm SFR_{\rm Pa\alpha}$. Their work spans a larger stellar mass range that does not reach the massive starbursts represented in our sample (Fig.\,\ref{Fig:NeufeldComp}). Consequently, \cite{Alberts_2026} infer $A_{\rm Pa\alpha}<0.2$ mag from H$\alpha$/Pa$\alpha$ ratios, which is also significantly less than what we find in our sample. The most massive starbursts at cosmic noon are significantly more attenuated than their main-sequence counterparts \citep{daCunha_2008, Hodge_2013, Casey_2014, Whitaker_2017}. In this work we demonstrate that this holds into the near-IR where the attenuation can reach up to 4 mag of attenuation at Pa$\alpha$.
In local LIRGs well above the $z\sim0$ main-sequence, \cite{Inami_2018} find that Br$\beta$-derived SFRs agree well with the $L_{\rm IR}$ SFRs when using the opacity models from mid-IR continuum fitting, which similarly predict low attenuation into the mid-IR. Although, our sample probes more extreme, dust-obscured systems than local LIRGs \citep[e.g.,][]{Kirkpatrick_2017,McKinney2023}.  
Compared to our $z\sim1-2$ sample, the SFR tracer differences may arise from redshift evolution in galaxy gas mass fractions and dust-to-stellar mass fractions, whereby for fixed stellar mass star-forming galaxies with higher gas fractions increase the birth-cloud/mixed dust contrast \citep{Santini2014, Casey2026}.


Figure~\ref{Fig:SF_MS} shows the position of our sample relative to the star-formation main sequence (SFMS) for each SFR diagnostic after correcting for attenuation. Since the galaxies in our sample are among the most luminous dust-obscured galaxies at $z\sim1-2$, we expect them to fall above the SFMS \citep[e.g.,][]{Yan2004,Kirkpatrick_2015,Kirkpatrick_2017}. Indeed, the 8 galaxies in the parent LRS sample with half-light radii from NIRCam \citep{DAWN_JWST_ARCHIVE_2025} have SFR surface densities ($\Sigma \rm SFR\equiv SFR/2\pi R_e^2$) above the expected main-sequence range by $\sim0.5$ dex \citep{Nadolny_2025}. All sources where the attenuation is inferred directly from multiple recombination lines fall above the SFMS. 

Some of the galaxies without multiple nebular lines, for which the average attenuation of the sample was assumed, 
overlap with the main sequence scatter (\textit{First} and \textit{Second} panel of Fig.~\ref{Fig:SF_MS}). The attenuation we assume in these source could plausibly underestimate the true attenuation, and therefore underestimate the dust-corrected SFR. Additionally, the Br$\alpha$ sources, which are the least sensitive to dust attenuation, occupy the lower-redshift end of our sample, where the effects of dust heating by old stars may be more pronounced. We do not interpret the location of the Br$\alpha$ sources on the SFMS as evidence that the Br$\alpha$ SFRs are intrinsically more accurate, particularly because they show the largest $\rm SFR_{line}$ and $\rm SFR_{IR}$ discrepancies (Figure~\ref{Fig:BrA_vs_PaA}). With the aforementioned caveats in mind, this could also be attributed to a past ($>10$ Myr) burst  boosting the $\sim 100$ Myr-averaged SFR$_{\rm IR}$ while the $\sim 10$ Myr-averaged SFR from Br$\alpha$ traces the time post-burst, reflecting either a declining or constant star-formation history. Medium band photometry or NIRSpec/PRISM spectra could be used to disentangle the competing effects of old stellar heating and declining star-formation histories as in \cite{Cooper_2024}. 


\section{Conclusions} \label{Conclusions}
In this work we select a sample of Pa$\alpha$, Br$\alpha$, and Br$\beta$-detected galaxies from an original sample of 37 LIRGs at $z\sim1-2$ with spectra from JWST MIRI/LRS and Spitzer IRS. We then compare dust corrected hydrogen recombination line SFRs with their corresponding $L_{\rm IR}$ SFRs to study the effects of dust attenuation in the near-IR and SFR indicators out to $z\sim2$. Our main conclusions are as follows:
\begin{itemize}
    \item Using multiple Paschen and Brackett emission lines, we derive attenuation between $1.87-4.05\,\mu$m which can be as high as $A_\lambda=4$ mag at $1.87\,\mu$m. We find that the IR continuum attenuation derived from \cite{McKinney_2025} is insufficient to correct for the amount of nebular attenuation that we measure, consistent with the differential reddening between optical continuum and optical nebular lines found in other surveys of optical recombination lines.
    \item The dust-corrected near-IR recombination-line SFRs are systematically lower than $\rm SFR_{\rm IR}$ with an offset of a factor of $\sim2.5$ on-average and up to an order of magnitude. We consider a variety of potential explanations including clumpy star-formation morphologies, declining or constant star-formation histories with a past burst, under-estimated nebular reddening, and biased $L_{\rm IR}$ SFRs from AGN and/or old stars. We disfavor the AGN heating scenario due to our ability to measure and subtract the AGN contribution to far-IR dust emission.
    \item We present the first sample of Br$\alpha$-derived SFRs in $z\sim1$ LIRGs,  with varying degrees of AGN contribution constrained by mid-IR spectroscopy out to $\sim24\,\mu$m. Relative to Pa$\alpha$ and Br$\beta$, the Br$\alpha$ SFRs exhibit the greatest discrepancy with IR-derived SFRs while being the least sensitive to dust.
\end{itemize}
 
It is evident that in the most massive, dust-obscured galaxies around cosmic noon the effects of dust attenuation are significant out to $\lambda_{\rm rest}\sim4\,\mu$m. Future observations using JWST NIRCam/grism and/or NIRSpec PRISM are needed to definitely characterize near-IR nebular line SFRs in dust-obscured cosmic noon starbursts. In particular, resolved Pa$\alpha$ morphologies could constrain the effects of clumpy morphologies, while PRISM spectroscopy would provide a handle on dust heating from evolved stellar populations via the Balmer Break strength, and greatly improved star-formation history modeling. 

\begin{acknowledgments}
The data were obtained from the Mikulski Archive for Space Telescopes at the Space Telescope Science Institute, which is operated by the Association of Universities for Research in Astronomy, Inc., under NASA contract NAS 5-03127 for JWST. These observations are associated with program \#3224. Support for program \#3224 was provided by NASA through a grant from the Space Telescope Science Institute, which is operated by the Association of Universities for Research in Astronomy, Inc., under NASA contract NAS 5-03127.

V.V. acknowledges support for the Galaxy Evolution Vertically Integrated Project (GEVIP) undergraduate research group from the National Science Foundation under NSF-AAG-1908817 and NSF-AAG-2009905.

V.V. would like to thank the mentorship of J.M., S.F, and A.T. during her undergraduate career as well as the astronomy department at the University of Texas at Austin for the great amounts of support they provide to undergraduates interested in research. V.V. would also like to thank her husband, her parents, and her family for their constant support.

\end{acknowledgments}

\begin{contribution}
V.V. was responsible for the formal analysis, investigation, methodology, software, validation, visualization, writing the original draft, editing, and implementing suggested comments.

J.M. came up with the initial research concept, provided supervision and mentorship throughout the project, and reviewed and edited the draft throughout the writing process.

A.P. and A.S. were responsible for the conceptualization and expansion of the research as well as providing meaningful feedback.

All other authors provided helpful feedback.

\end{contribution}

\begin{deluxetable*}{llrlllll}
\tablecaption{Recombination Line Luminosities and Derived Star Formation Rates. $f_{\rm AGN}$ denotes the AGN fractional contribution to the mid-IR. Stellar masses are derived from SED fitting by CIGALE. SFRs are measured using \texttt{emcee} as described in \ref{sec:linemodels} with units of $\rm M_\odot\,yr^{-1}$, and have been corrected for dust attenuation as described in Section \ref{sec:dustAttenuation}. \label{tab:recombination}}
\tablehead{
\colhead{ID} & \colhead{$z_{\rm spec}$} & \colhead{$f_{\rm AGN}$ (\%)} & \colhead{$M_\ast$ ($M_\odot$)} & \colhead{$\rm SFR_{Pa\alpha}$} & \colhead{$\rm SFR_{Br\beta}$} & \colhead{$\rm SFR_{Br\alpha}$} & \colhead{$\rm SFR_{IR}$}
}
\startdata
FLS-IRS-8493 & 0.647 & 35 & 10.2 $\pm 0.2$ & -- & -- & $7_{-1}^{+1}$ & $854 \pm 393$ \\
FLS-IRS-8040 & 0.753 & 38 & 10.3 $\pm 0.2$ & -- & $<1460$ & $32_{-6}^{+5}$ & $459 \pm 211$ \\
GN-IRS-27 & 0.776 & 35 & 11.2 $\pm 0.2$ & $<159$ & $<1194$ & $16_{-3}^{+3}$ & $183 \pm 84$ \\
GN-IRS-50 & 0.822 & 0 & 10.7 $\pm 0.2$ & -- & -- & $14_{-3}^{+3}$ & $71 \pm 33$ \\
GN-IRS-61 & 0.842 & 0 & 10.8 $\pm 0.2$ & -- & -- & $11_{-2}^{+3}$ & $82 \pm 38$ \\
GN-IRS-42 & 0.971 & 18 & 10.6 $\pm 0.2$ & -- & $<2990$ & $29_{-5}^{+6}$ & $214 \pm 99$ \\
GN-IRS-6 & 1.009 & 0 & 10.8 $\pm 0.2$ & -- & -- & $<922$ & $126 \pm 58$ \\
GN-IRS-5 & 1.150 & 0 & 10.9 $\pm 0.2$ & -- & $73_{-25}^{+25}$ & $<1244$ & $270 \pm 124$ \\
GN-IRS-26 & 1.220 & 2 & 11.5 $\pm 0.2$ & -- & $156_{-29}^{+31}$ & $<1701$ & $663 \pm 305$ \\
GN-IRS-2 & 1.242 & 0 & 10.8 $\pm 0.2$ & -- & $<2084$ & $68_{-12}^{+13}$ & $214 \pm 99$ \\
GN-IRS-4 & 1.264 & 16 & 10.8 $\pm 0.2$ & -- & $<1153$ & $<1049$ & $340 \pm 157$ \\
GN-IRS-1 & 1.436 & 19 & 10.7 $\pm 0.2$ & -- & $229_{-44}^{+47}$ & $<1017$ & $409 \pm 188$ \\
GN-IRS-63 & 1.466 & 9 & 10.9 $\pm 0.2$ & -- & $112_{-20}^{+22}$ & $80_{-14}^{+13}$ & $196 \pm 90$ \\
GN-IRS-3 & 1.515 & 0 & 11.1 $\pm 0.2$ & -- & $<3898$ & $152_{-27}^{+31}$ & $356 \pm 164$ \\
FLS-IRS-22722 & 1.802 & 77 & 10.1 $\pm 0.2$ & $132_{-67}^{+18}$ & -- & -- & $633 \pm 292$ \\
FLS-IRS-289 & 1.844 & 37 & 10.9 $\pm 0.2$ & $386_{-43}^{+45}$ & $<7517$ & -- & $2002 \pm 922$ \\
GN-IRS-19 & 1.886 & 63 & 11.6 $\pm 0.2$ & $364_{-40}^{+44}$ & $<644$ & -- & $418 \pm 193$ \\
GN-IRS-15 & 1.962 & 22 & 11.2 $\pm 0.2$ & $637_{-66}^{+72}$ & -- & -- & $1152 \pm 531$ \\
FLS-IRS-22530 & 1.963 & 21 & 10.9 $\pm 0.2$ & $741_{-74}^{+74}$ & $741_{-132}^{+151}$ & -- & $1784 \pm 822$ \\
GN-IRS-21 & 1.974 & 47 & 11.3 $\pm 0.2$ & $417_{-50}^{+45}$ & $377_{-69}^{+72}$ & $<2063$ & $936 \pm 431$ \\
GN-IRS-11 & 1.989 & 23 & 11.6 $\pm 0.2$ & $214_{-68}^{+52}$ & $<1224$ & $<3516$ & $577 \pm 266$ \\
GN-IRS-7 & 1.998 & 0 & 11.0 $\pm 0.2$ & $526_{-220}^{+59}$ & $<2397$ & $<5496$ & $564 \pm 260$ \\
GN-IRS-58 & 1.999 & 32 & 10.9 $\pm 0.2$ & $81_{-9}^{+10}$ & -- & -- & $289 \pm 133$ \\
GN-IRS-12 & 2.009 & 79 & 10.7 $\pm 0.2$ & $179_{-24}^{+22}$ & $168_{-40}^{+46}$ & $<3373$ & $980 \pm 452$ \\
FLS-IRS-521 & 2.017 & 40 & 11.0 $\pm 0.2$ & $792_{-74}^{+80}$ & -- & -- & $1665 \pm 767$ \\
FLS-IRS-8226 & 2.119 & 69 & 11.4 $\pm 0.2$ & $388_{-144}^{+48}$ & $475_{-85}^{+110}$ & $<2987$ & $1027 \pm 473$ \\
FLS-IRS-509 & 2.177 & 77 & 10.5 $\pm 0.2$ & $546_{-185}^{+75}$ & $<830$ & $<729$ & $428 \pm 197$ \\
\enddata
\end{deluxetable*}

\clearpage

\bibliography{Bibliography}{}
\bibliographystyle{aasjournalv7}
\nocite{*}

\end{document}